\documentclass[twocolumn]{aastex63}

\usepackage{lineno}

\usepackage{booktabs}
\usepackage{amsmath}

\newcommand{\psr}{J1846$-$0258}
\newcommand{\pwn}{PWN~Kes~75}
\newcommand{\fermi}{\textit{Fermi}-LAT}

\graphicspath{{./}{gfx/}}

\shorttitle{\fermi\ observations of PWN Kes 75}
\shortauthors{Straal et al.}

\begin{document}

\title{Discovery of GeV gamma-ray emission from PWN Kes~75 and PSR~J1846$-$0258}

\correspondingauthor{Samayra M. Straal}
\email{samayrastraal@gmail.com}

\author[0000-0003-4136-7848]{Samayra~M.~Straal}
\affil{NYU Abu Dhabi, PO Box 129188, Abu Dhabi, United Arab Emirates}
\affiliation{Center for Astro, Particle, and Planetary Physics (CAP$^3$), NYU Abu Dhabi, PO Box 129188, Abu Dhabi, United Arab Emirates}

\author[0000-0003-4679-1058]{Joseph~D. Gelfand}
\affil{NYU Abu Dhabi, PO Box 129188, Abu Dhabi, United Arab Emirates}
\affiliation{Center for Astro, Particle, and Planetary Physics (CAP$^3$), NYU Abu Dhabi, PO Box 129188, Abu Dhabi, United Arab Emirates}
\affiliation{Center for Cosmology and Particle Physics (CCPP, Affiliate), New York University, 726 Broadway, Room 958, New York, NY 10003}

\author{Jordan L. Eagle}
\affil{Clemson University, 105 Sikes Hall, Clemson, SC 29634}
\affiliation{Harvard \& Smithsonian Center for Astrophysics, 60 Garden St, Cambridge, MA 02138}

\begin{abstract}
We report the detection of gamma-ray emission from \pwn\ and PSR \psr.
Through modeling the spectral energy distribution incorporating the new \fermi\ data, we find the the observed gamma-ray emission is likely a combination of both the PWN and pulsar magnetosphere. 
The spectral shape of this magnetospheric emission is similar to the $\gamma$-ray spectrum of rotation powered pulsars detected by \fermi\ and the results from our best-fit model suggest the pulsar's magnetospheric emission accounts for 1\% of the current spin-down luminosity. 
Prior works attempted to characterize the properties of this system and found a low supernova explosion energy and low SN ejecta mass. 
We re-analyze the broadband emission incorporating the new \textit{Fermi} emission and compare the implications of our results to prior reports. 
The best-fit gamma-ray emission model suggests a second very hot photon field possibly generated by the stellar wind of a Wolf-Rayet star embedded within the nebula, which supports the low ejecta mass found for the progenitor in prior reports and here in the scenario of binary mass transfer.

\vspace{2cm}
\end{abstract}

\section{Introduction}
\label{sec:introduction}
Neutron stars are formed in core-collapse supernovae (SN) and have a wide range of observational manifestations, such as rotation-powered pulsars (RPPs), central compact objects (CCOs), and magnetars.
The relationship between these manifestations is unknown, and they could be seperate objects or different phases of evolution \citep[see e.g.,][for a review]{harding2013,kaspi2018}.
The vast majority of the over 3000 known neutron stars are observed as RPPs, where the associated emission comes from the radiating relativistic particles and is powered by the loss of the neutron star's rotational energy.
A less common class of neutron stars are magnetars, for whom the bursting activity and, in nearly all cases, requires an energy source larger than their available rotational energy.  
In these sources, the energy released by the decay of extremely strong ($\gtrsim10^{14}~{\rm G}$) surface magnetic field is believed to the dominant contributor to their observed emission \citep{kaspi2017}.
Unlike RPPs, magnetars produce outbursts in X-rays and gamma-rays and are generally not observed to show radio pulsations.
In the past decade, the distinct border between RPPs and magnetars have started to fade with some rotation powered pulsars having a similarly strong spin-down inferred dipolar surface magnetic field and the discovery of magnetar-like outbursts in a couple of young pulsars. 
This fading distinction is also present in the P-Pdot diagram for pulsars, challenging the source class distinction between magnetars and RPPs.
Amongst the previously assumed `regular' RPPs displaying magnetar-like behavior is the X-ray pulsar \psr\,, powering the pulsar wind nebula (PWN) Kes~75 \citep{gavriil2008}.  
With a surface dipolar magnetic field of $B = 5\times10^{13}$\,G, a spin-down luminosity of $\dot{E} = 8.1\times10^{36}$\,erg\,s$^{-1}$, and a characteristic age of $\tau_c = 723$\,yr, \citep{gotthelf2000,livingstone11} this pulsar was believed to be a RPP with a high surface-magnetic field.

In 2006 this pulsar was detected to emit magnetar-like bursts \citep{gavriil2008} and its spectrum observed to change with time \citep{kumar2008,ng2008}.  
After the observed variability the pulsar returned to its previous (quiescent) state, with the exception of its braking index that was now measured to be $n = 2.16\pm0.13$ \citep{livingstone11} rather than the $n=2.65$ it was before the outburst. 
In recent years, PSR \psr\ showed another period of magnetar-like bursts \citep{krimm2020,blumer2021}, before returning to `normal' X-ray luminosities. 

The behavior of PSR \psr\ indicates that perhaps young pulsars can either evolve into RPPs or the magnetar class by an emerging surface magnetic field (e.g. \citealt{gullon15}).  
In order to understand why some pulsars show magnetar-like behavior, it is crucial to understand the circumstances that lead to the formation of the neutron star (NS).  
Hence, determining the properties of the NS at birth as well as its progenitor is key.  

If the rotational power output is strong enough from the pulsar, a PWN can form.
A PWN is a highly relativistic plasma consisting of electrons and positrons that were injected by the central pulsar. 
The interaction between the colder pulsar wind and the relativistic plasma generates a shock wave called the "termination" shock. 
The particles are injected by the central pulsar and accelerated at the termination shock as they enter the nebula.
Their evolution strongly depends on the SN and NS characteristics \citep{gaensler06}.
By studying the spectral and dynamical evolution of a PWN, combined with the observed properties of the pulsar, one can then obtain information on the surroundings of the SNR, the NS birth properties, such as its initial spin period and spin down energy 
\citep[see e.g.,][]{gelfand09}. 
Applying this type of study to PSR \psr\ and \pwn\ allows valuable constraints in understanding why this source exhibits both RPP-like and magnetar properties.  
From this we can understand whether this is an evolutionary phase of young and energetic NSs, or a separate class.

\pwn\ has been the focus of multiple studies from radio to infra-red, X-rays, and TeVs \citep[see e.g.,][]{salter89a,reynolds18,HGPS,temim2019,gotthelf2021}.
Kes 75 radiates via synchrotron emission at lower energies and via Inverse Compton Scattering (ICS, ) at TeV gamma-ray energies. 
However, for precise characterization of the IC spectrum and thus local photon field characterization, MeV--GeV measurements are imperative.
  
The properties of the photon fields 
such as the temperature and energy density, enable a more complete picture of the local environment for Kes 75, which in turn constrains the magnetic field of the PWN.
Previous attempts in the search for gamma-ray emission coincident with Kes 75 have been unsuccessful. 
However, a 4-$\sigma$ \fermi\ detection of the pulsar's pulsed $\gamma$-ray emission up to 100\,MeV, has been reported by \citet{kuiper2018}.
This work however, is focused on energies above 100\,MeV, where the pulsed emission is thus expected to not be significant.

In this paper, we report the likely detection of the PWN to the Kes 75 complex with the \fermi. 
We analyze the contributions from both the pulsar and nebula to characterize the observed MeV--GeV emission. 
We first discuss the data analysis and the results in Sect. \ref{sec:Data}. 
We use these results in modeling of the PWN in Sect. \ref{sec:pwnmodel} and discuss our findings and their implications in Sect. \ref{sec:discussion} before we conclude in Sect. \ref{sec:conclusions}.

\section{\fermi\ Analysis and Results} 
\label{sec:Data}
The \textit{Fermi}-LAT (Large Area Telescope) is a pair conversion telescope detecting $\gamma$-ray photons between 20\,MeV to more than 1\,TeV.  

Since beginning operation in 2008 August, the telescope has performed all-sky surveys every 3 hours. 
The recently improved sensitivity and spatial resolution of the instrument is enabled by the Pass 8 update \citet{atwood2013}.
The Pass 8 update, in its current version P8R3 \citep{bruel2018} offers a higher acceptance of detected photons and a narrower point spread function (PSF) at higher energies, allowing for better analysis of point sources at higher energies.  
A main addition to the update was the inclusion of PSF quartiles.
The \fermi\ detected photons are divided into four quartiles,
categorized based on their reconstruction accuracy where PSF 0 has the lowest quality and PSF3 has the highest quality\footnote{\url{https://fermi.gsfc.nasa.gov/ssc/data/analysis/documentation/Cicerone/Cicerone_LAT_IRFs/IRF_PSF.html}}.
This enables the observer to specify the PSF quality of the photons, to optimize between spatial resolution and sensitivity.

\subsection{Data selection}
\label{sec:data_selection}
We have analyzed $11.5$~years (2008 August 4 to 2020 February 26) of data towards \pwn\, selecting data in a 20 degree radius centered on the X-ray coordinates of \pwn\,, RA: 18$^{\rm h}$46$^{\rm m}$25$^{\rm s}$ and declination: -02\degr59\arcmin13\arcsec. 
We constrained our data selection to the 100\,MeV $-$ 500\,GeV energy range. 
We allowed a maximum zenith angle of 100 degree to reduce earth limb contamination.
The data was selected to only include photons in the upper three quartile selection of point-spread functions (PSFs 3, 2, 1) and binned to 0.1$^\circ$ per pixel to accommodate the resolution at higher energies, with 8 bins per decade in energy.  
We consider sources up to a 10$^\circ$ radius in the global source model.   
To model the Galactic diffuse background we use template gill$\_$iem$\_$v07.fits and for the isotropic diffuse backgrounds the iso$\_$P8R3$\_$SOURCE$\_$V3$\_$v1.txt template.  
As the analysis is performed using more than 10 years of data,  we used the 4FGL-DR3 catalog \citep{4FGLDR3}, version gll$\_$psc$\_$v27.fit to build our source model.
The analysis is performed using fermitools v.2.2.0, with fermipy v1.1.6 \citep{wood2017}. 

\subsection{Data analysis}
\label{sec:data_analysis}
To investigate if there is a detection of \pwn\ and pulsar \psr\ we consider nearby unassociated\footnote{sources not associated with any astronomical counterpart} sources to be a possible counterpart.
The nearest source within this region, at 0.230$^\circ$, was 4FGL~J1846.9-0247c, described as a point source with a LogParabola spectrum.  Its location and spectrum make it a good candidate to actually describe emission from \pwn\ and hence we exclude this source from the model.
In the new catalog, 4FGLDR3 \citep{4FGLDR3}, a new source is reported near the location of \pwn\ , 4FGL~J1845.8-0236c, at 0.4 degree, significantly further away. We do not consider this source to be the counterpart.
Indeed, we achieve the best-fit model where J1846.9-0247c is replaced by a point source at the position of \pwn\ and includes the nearby, unassociated source J1845.8-0236c, which is modeling excess diffuse emission unrelated to \pwn.

We then fit for a point source at the location of \pwn\ and compute a Test-Statistic (TS) map with the abovementioned source excluded.
The TS is defined to be the natural logarithm of the difference in the likelihood of one hypothesis (e.g. presence of one additional source) and the likelihood for the null hypothesis (e.g. absence of source):
\begin{equation}
    TS = 2 \times \frac{\log{\mathcal{L}_{1}}}{\log{\mathcal{L}_{0}}}.
\end{equation}
The TS quantifies how significantly a source is detected with a given set of location and spectral parameters and the significance of such a detection can be estimated by taking the square root of the TS.
For a significant detection, the \fermi\ detection threshold is a TS = 25 for 4 DOF (degrees of freedom).
As shown in Figure \ref{fig:tsmap}, there is significant residual TS (TS$\_$max $\sim$ 70, N$\_$pred = 2112) from a point source coincident with Kes 75 for energies from 100\,MeV -- 500\,GeV, fitted with a powerlaw spectrum with index = 2.0.
Allowing the point source to vary in location did not improve the fit.

A previous examination of this region using the 4FGLDR2 model in which the newly reported source 4FGL~J1845.8-0236c was not included showed an excess in that region. After inclusion of this source in 4FGLDR3 no such excess is observed in the analysis presented in this work. 
Given its proximity to \pwn\ we tried to optimize the location for both \pwn\ and the new source by running a localization analysis on both sources in fermipy. 
No better location was found within 0.5 degree of each source.

Even though \pwn\ is much smaller than the PSF of the \fermi\ instruments ($\sim$0.15\degr\ for E $>$ 10 GeV)\footnote{\url{https://fermi.gsfc.nasa.gov/science/instruments/table1-1.html}}, we have tested the source for extension using the extension templates RadialDisk and RadialGaussian. 
Using the radial disk template we find a best-fit extension of $0.283^{+0.021}_{-0.033}$ degree and a 95$\%$ upper limit of $0.32$ degree with $TS_{\rm ext} = 12.5$. 
For the radial gaussian template, we find a size of $0.247^{+0.053}_{0.048}$, and a 95$\%$ upper limit of $0.34$ degree with $TS_{\rm ext} = 15.0$. 
Both these extensions are not significant and we conclude there is no evidence for source extension. 
We additionally tested the extension results by changing the Galactic diffuse model flux by $\pm 5\%$ (as we do for the measured flux, see below), and find significant changes in the extension results.
Because \pwn\ lies in a complex region of diffuse emission, we therefore interpret the extension results as not significant and favor instead the point-source model.

To investigate the systematic uncertainties in the Galactic diffuse background and effective area, we analyzed the data using the following configurations: 
\begin{itemize}
\item the 4FGL catalog in stead of 4FGL-DR3, 
\item the previous version of the Pass 8 instrument response function (IRF) newly released isotropic diffuse background version P8R3$\_$SOURCE$\_$V2, and
\item by reducing the maximum zenith angle to 90 degrees, further decreasing any earth limb contamination.
\item by altering the normalization of the diffuse background by $\pm 5\%$.
\end{itemize}

The obtained spectra as a result of using different configurations in the data analysis are shown in Figure \ref{fig:flux} and show good agreement with each other.
When applying the 4FGL catalog, we included the newly reported source 4FGL~J1845.8-0236c to remain consistent with the other source models. 
For the remainder of this report we adopt the maximum and minimum bounds (including upper limits) from the tested configurations to represent the systematic error (see Table \ref{tab:model_obs}). 

The obtained spectrum describes the entire PWN complex, which is both the PWN, and the pulsar.
In Sec. \ref{sec:pwnmodel} we find the energies where the pulsar dominates ($100$\,MeV $ < E < 5$\,GeV) and where the PWN dominates ($5$\,GeV $ < E < 500$\,GeV).
We therefore also test spectral models best describing both components in the \fermi\ source model.
The spectrum in energies $100$\,MeV $ < E < 5$\,GeV is best described using a PLEC2 model (powerlaw with super exponential cutoff 2\footnote{\url{https://fermi.gsfc.nasa.gov/ssc/data/analysis/scitools/source_models.html##PLSuperExpCutoff2}}, and we find an improvement of the fit compared to fitting a simple powerlaw. The TS for \pwn\ in this energy range also increases from 73 to 89. 
Its best fit parameters have an index1 = $1.46\pm0.44$, index2 = $0.67$, the exponent factor = $0.009\pm2.6\times10^{-6}$ and a prefactor of $(7.5\pm4.5)\times10^{-12}$.
The data above 5\,GeV is best described by a powerlaw with index = $2.49\pm0.38$ and prefactor = $(2.32\pm2.25)\times10^{-12}$, with a TS of 16.5.
Further in this work we model the full spectrum of the PWN complex, using the obtained fluxes given in Table \ref{tab:model_obs}.

We note that the obtained spectrum of \pwn\ is very similar to the unassociated source 4FGL~J1846.9-0247c located at 0.230 degrees from the X-ray coordinates of \pwn. 
Given the obtained source spectrum (see Sec. \ref{sec:spectral}), the detection of \pwn\ at energies of 0.332\,TeV and above in the HESS Galactic plane survey \citep{HGPS} makes us conclude that the unassociated source 4FGL~J1846.9-0247c describes emission fom \pwn.

\begin{figure*}[ht!]
    \centering
    \gridline{\fig{20221011_4FGLDR3_TSmap.pdf}{0.49\textwidth}{}
        \fig{20221011_4FGLDR3_TSmap_1GeV_Above.pdf}{0.49\textwidth}{}
    }
    \caption{TS maps centered at the location of \pwn. These maps show the TS of an additional point source at any pixel on the map. The image on the left shows the TS map for energies between 100\,MeV and 500\,GeV and the right for energies between 1\,GeV and 500\,GeV. The inner contour in the right figure is drawn at TS $=40$.}\label{fig:tsmap} 
\end{figure*}

\begin{figure}[ht!]
    \centering
    \includegraphics[width=0.99\columnwidth]{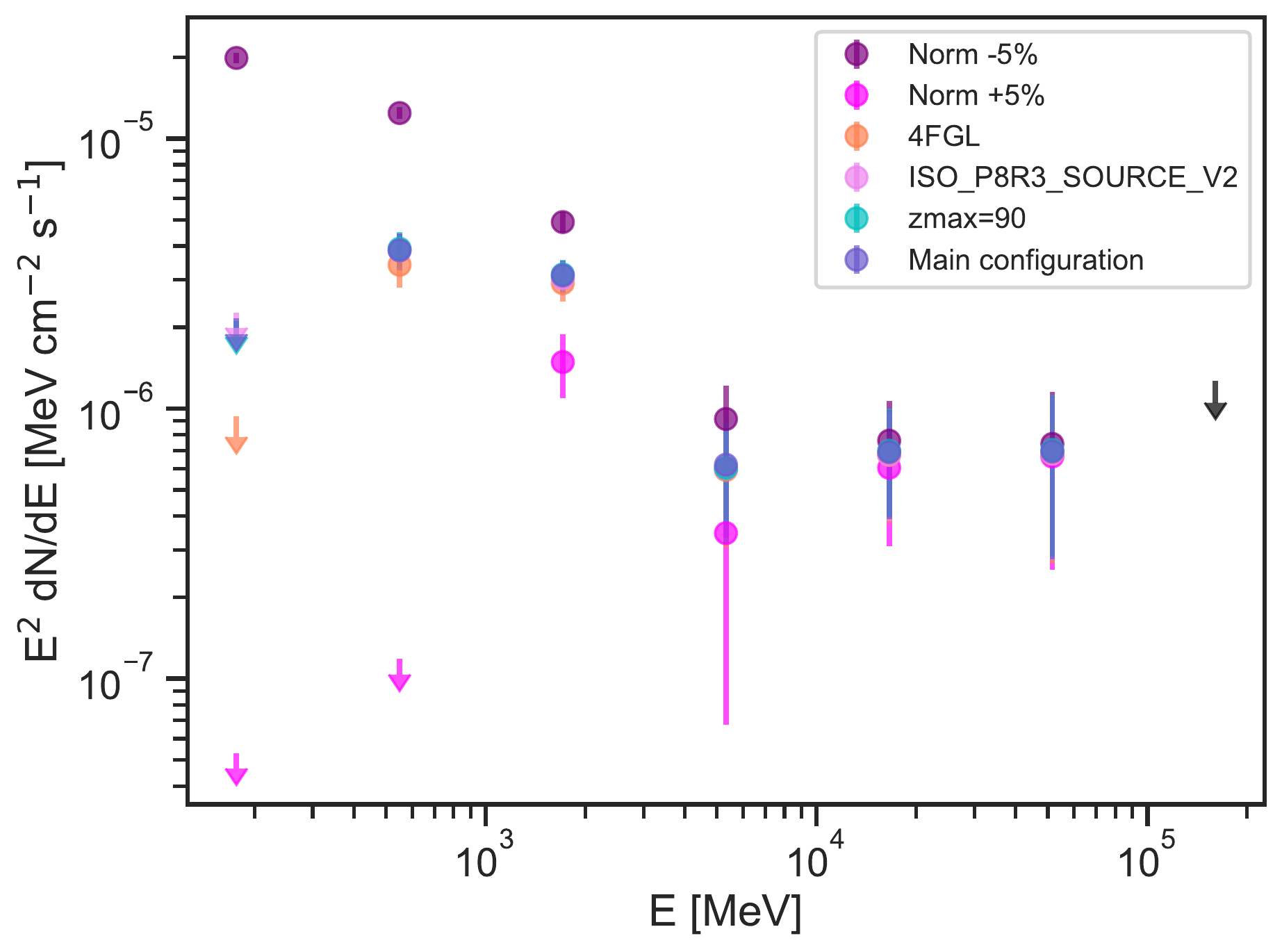}
    \caption{Fluxes of a point source at the location of \pwn, using six different configuration files to assess the influence on the systematic errors. See Sec. \ref{sec:data_analysis} for the different configurations.}
    \label{fig:flux}
\end{figure}

\section{PWN Modeling}
\subsection{Spectral Analysis}
\label{sec:spectral}
The fluxes obtained above and given in Table \ref{tab:model_obs} are for fitting a $\gamma$-ray point source at the X-ray coordinates of \pwn.
Furthermore, since the PWN complex is unresolved by \fermi\, this source contains emission from the PWN and the pulsar.
Due to the lack of non-thermal X-ray emission observed from the supernova remnant (SNR) shell \citep{gotthelf2021}, it does not seem likely to contribute significantly.
Given the previous non-detection of pulsed gamma-ray emission in the energy band analyzed in this work \citep{kuiper2018}, we assume magnetospheric pulsar emission typical to other gamma-ray pulsars, which not necessarily arises from the same emission mechanism responsible for the detection at lower energies ($<$ 100\,MeV).
The $\gamma$-ray emission from typical gamma-ray pulsars is primarily observed below $\lesssim1-10~{\rm GeV}$, and well described by a power-law or power-law with an exponential cutoff \citep{2fpc}. 
As described below, we model the observed $\gamma$-ray emission of this source with a power-law with an exponential cutoff typical of a pulsar and the $\gamma$-ray emission from the PWN as predicted by our model described below.

\subsection{PWN Model}
\label{sec:pwnmodel}
As the evolution of a PWN inside a SNR depends heavily on the properties of the progenitor star, supernova explosion, and the neutron star at birth, modeling the PWN allows for obtaining information on the aforementioned. 
Following both the dynamical and spectral evolution of the PWN is currently best done using one-zone models.
We use the evolutionary model as described by \citet{gelfand09} to obtain values for the properties given in Table \ref{tab:model_pars}. 
The procedure used to fit the observed properties of a PWN with the values predicted by this model for a particular combination of input parameters is described in detail by \citet{hattori2020}, and we refer the reader to those papers for an extended description of the model and its applications.  
We do note this model and fitting procedures has successfully reproduced the properties of several PWNe, all with different sets of measured properties. \citep[e.g.,][]{hattori2020,gotthelf14,gelfand15,gelfand13}.

The properties we aim to reproduce with this model are listed in Table \ref{tab:model_obs}, 
and include the \fermi\ flux measurements derived in \S \ref{sec:data_analysis}.
Adding these flux measurements to the existing spectral energy distribution (SED) of \pwn\ allows us to disentangle the PWN and the possible pulsar component in the \fermi\ detection.
For the pulsar we assume a flux contribution in GeV emission that can be described by a power-law with exponential cut off
\begin{equation}
    \frac{dN_{\gamma}}{dE} = N_0 \times E^{-\Gamma} \times \rm{exp}\left(\frac{-E}{E_{\rm cut}}\right),
\end{equation}
that gives the pulsar flux in ergs cm\,$^{-1}$\,s$^{-1}$, $N_0$ the normalization at 1\,GeV in photons\,s$^{-1}$\,cm$^{-1}$\,GeV$^{-1}$ and the energy E is given in GeV, $\Gamma$ the photon index, and E$_{\rm cut}$ the cut off energy in GeV.
For consistency with previous analyses of this source \citep{gelfand2014,reynolds18,gotthelf2021}, we assume the pulsar braking index is a constant $p = 2.65$ \citep{livingstone11}, the value measured before the outbursts detected in 2009, throughout its evolution.

From our modeling of the pulsar component, we obtain a gamma-ray efficiency of $0.97\%$, a cutoff energy of $\sim 1$\,GeV, and a spectral index of $\Gamma = 1.29$ (see Table \ref{tab:model_pars}). 

\begin{figure}[ht!]
    \centering
    \includegraphics[width=0.99\columnwidth]{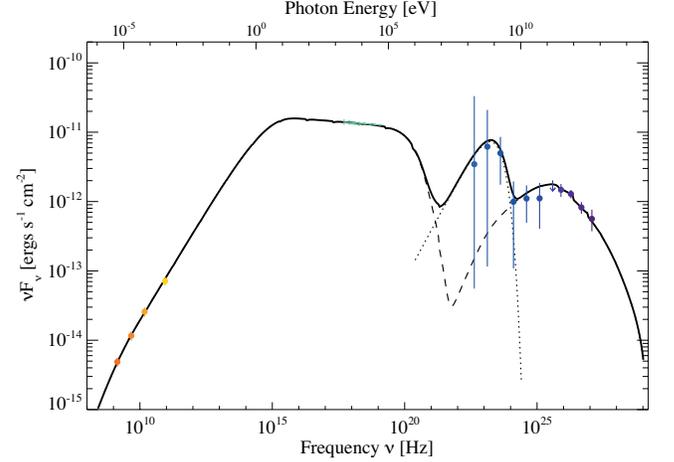}
    \caption{SED of \pwn. Observed data are shown in color and best model fit in black solid line. Dashed and dotted line respectively show the contribution of the PWN and pulsar. Observed and best-fit properties can be found in Tables \ref{tab:model_obs} and \ref{tab:model_pars}. The blue \fermi\ data points reflect the fluxes obtained in the main configuration. The model is fit to the range of fluxes as described in Sec. \ref{sec:data_analysis}.}
    \label{fig:flux_sed}
\end{figure}

With four free parameters in our modeling, our best-fit model (see Fig. \ref{fig:flux_sed}) reaches a chi-squared ($\chi^2$) value of 2.4. 

The best fit model shows that the fermi energies below $\sim 5$\,GeV are described by the pulsar component of the model, wheras the energies above $\sim 5$\,GeV are described by the PWN component of our model.

The values of the input parameters for the best-fit model are given in Table \ref{tab:model_obs} and the modeling results will be further discussed in Sec. \ref{sec:disc_model}.

\begin{table*}[ht!]
    \caption{Observed properties of \pwn\ used in the modeling of this source}
    \label{tab:model_obs}
    \resizebox{0.95\linewidth}{!}{%
    \centering
    \begin{tabular}{cccc}
    \hline
    \hline 
    {\sc Property} & {\sc Observed} & {\sc Model} & {\sc Citation} \\
    \hline
    \multicolumn{4}{c}{\it PSR \psr } \\
    $\dot{E}(\frac{\rm erg}{\rm s})$ & $8.10\times10^{36}$ & Fixed & \citet{livingstone11}\\
    $t_{\rm ch}$ (yr) & 723 & Fixed & \citet{} \\
    $p$ & $2.65\pm{0.01}$ & Fixed & \\
   $P$ (ms) & 326.57 & - & \citet{livingstone11}\\

    \multicolumn{4}{c}{\it Pulsar Wind Nebula} \\
    Angular radius $\theta_{\rm pwn}$ & $30\arcsec \pm 1.7\arcsec$ & $30\farcs3$  & \citet{} \\
    Angular expansion rate $\dot{\theta}_{\rm pwn}$ & $(0.249\pm0.023)\frac{\%}{\rm year}$ & $0.253 \frac{\%}{\rm year}$ & \cite{reynolds18} \\
    $S_{1.4}$ (mJy) & $348\pm52$ & 341 &  \citealt{salter89a}\\
    $S_{4.7}$ (mJy) & $247\pm37$ & 254 & \citealt{salter89a} \\
    $S_{15}$ (mJy)  & $172\pm26$ & 168 & \citealt{salter89a} \\
    $S_{89}$ (mJy)  & $80\pm12$  & 86 & \citealt{bock05} \\
    $F$(2$-$10~keV)$^a$ & $2.031\pm{0.025} \times10^{-12}$ & $2.130\times10^{-12}$ & \citet{gotthelf2021}\\
    $\Gamma$(2$-$55~keV) & $2.13\pm0.022$ & $2.04$ & \citet{gotthelf2021} \\
    $F$(0.176 GeV)$^a$ & ($<8.49\times10^{-2}-31.84)^{1.3}\times10^{-12}$ & $4.78\times10^{-12}$ & This work \\
    $F$(0.549 GeV)$^a$ & ($<0.189-19.91)^{1.0}\times10^{-12}$ & $7.50\times10^{-12}$ & This work \\
    $F$(1.71 GeV)$^a$  & ($2.38-7.86)^{0.70}_{-0.63}\times10^{-12}$ & $5.61\times10^{-12}$ & This work \\
    $F$(5.32 GeV)$^a$  & ($0.554-1.46)^{0.48}_{-0.45}\times10^{-12}$   & $1.21\times10^{-12}$ & This work \\
    $F$(16.6 GeV)$^a$  & ($0.968-1.22)^{0.49}_{-0.47}\times10^{-12}$   & $1.35\times10^{-12}$ & This work \\
    $F$(51.6 GeV)$^a$  & ($1.07-1.18)^{0.67}_{-0.66}\times10^{-12}$   & $1.65\times10^{-12}$ & This work \\
    $F$(160 GeV)$^a$   & $<2.03\times10^{-12}(3\sigma)$ & $1.71\times10^{-12}$ & This work \\
    $F$(0.332 TeV)$^a$ & $8.38^{1.81}_{-1.78}\times10^{-12}$ & $7.72\times10^{-12}$ & \citet{HGPS} \\
    $F$(0.787 TeV)$^a$ & $1.30^{0.18}_{-0.18}\times10^{-12}$ & $1.20\times10^{-12}$ & \citet{HGPS} \\
    $F$(1.96 TeV)$^a$  & $1.34^{0.26}_{-0.25}\times10^{-13}$ & $1.41\times10^{-13}$ & \citet{HGPS} \\
    $F$(4.87 TeV)$^a$  & $1.48^{0.53}_{-0.50}\times10^{-14}$ & $1.30\times10^{-14}$ & \citet{HGPS} \\
    
    \multicolumn{4}{c}{\it Supernova Remnant} \\
    Angular radius $\theta_{\rm snr}$ & $1\farcm50 \pm 0\farcm15$ & $1\farcm44$ &  \citet{} \\
    Distance $d$ & ($5.8^{0.5}_{-0.4}$)~kpc & $5.82$~kpc & \citet{verbiest2012}\\
    \hline
    \hline
    \end{tabular}
    }
    \tablenotetext{a}{Unabsorbed flux given in units of ergs s$^{-1}$ cm$^{-2}$.}
\end{table*}

\begin{table*}[ht!]
    \caption{Derived properties of \pwn\ from modeling of this source}
    \label{tab:model_pars}
    \resizebox{0.95\linewidth}{!}{%
    \centering
    \begin{tabular}{ccc}
    \hline
    \hline 
    {\sc Property} & {\sc Derived in \citet{gotthelf2021}} & {\sc Derived in this work}  \\
    \hline
    Supernova Explosion Energy $E_{\rm sn}$ & $0.126\times10^{51}~{\rm ergs}$ & $0.11\times10^{51}~{\rm ergs}$ \\
    Supernova Ejecta Mass $M_{\rm ej}$ & 0.51~M$_\odot$ & 0.37~M$_\odot$ \\
    ISM Density $n_{\rm ism}$ & 0.56~cm$^{-3}$ & 0.63~cm$^{-3}$ \\
    Wind Magnetization $\eta_{\rm B}$ & $0.0724$ & $0.115$ \\
    Minimum Energy of Injected Leptons $E_{\rm min}$ & 2.0~GeV & 1.9~GeV \\
    Break Energy of Injected Leptons $E_{\rm break}$ & 2.042~TeV & 1.92~TeV  \\
    Maximum Energy of Injected Leptons $E_{\rm max}$ & 1.00~PeV & 1.11~PeV \\
    Low-Energy Particle Index $p_1$ & 1.73 & 1.17 \\
    High-Energy Particle Index $p_2$ & 3.04 & 3.05 \\
    Temperature of External Photon Field 1 $T_{\rm ic}$ & 32~K & 32~K \\
    Normalization of External Photon Field 1 $K_{\rm ic}$ & $1.17\times10^{-3}$ & $2.88\times10^{-3}$\\
    Temperature of External Photon Field 2 $T_{\rm ic}$ $^a$ & $-$ & $1.46\times10^{5}$~K \\
    Normalization of External Photon Field 2$K_{\rm ic}$ $^a$ & $-$ & $1.40\times10^{-15}$\\
    
    \multicolumn{3}{c}{\it Pulsar} \\    
    Pulsar Spindown Timescale $\tau_{\rm sd}$ & 398~years & 432~years \\
    Pulsar Initial Spin-down Luminosity $\dot{E}_0$ & $4.69\times10^{37}$\,ergs\,s$^{-1}$ & $3.9\times10^{37}$\,ergs\,s$^{-1}$ \\
    Pulsar Initial Spin Period $P_0$ & $\approx 200$\,ms & $\approx268~{\rm ms}$  \\
    Pulsar photon index ($\gamma$-rays) $^a$ & $-$ & 1.29\\
    Pulsar cutoff Energy E$_{\rm cut}$ $^a$ & $-$ & $\sim 1$\,GeV\\
    Gamma-ray efficiency $\eta = L_{\gamma}/\dot{E}$ $^a$ & $-$ & 0.97\% \\
    Gamma-ray Luminosity $L_{\gamma}$ ($0.1-100$\,GeV) & $-$ & $7.86\times10^{34}$\,ergs\,s$^{-1}$\\
    \hline
    $\chi^2$ / degrees of freedom & $-$ &2.4 / 4  \\
    \hline
    \hline
    \end{tabular}
    }
    \tablenotetext{a}{Parameters not modeled in \citet{gotthelf2021}.}
\tablecomments{The free parameters in the physical models used to reproduce the observed properties of \pwn\ are listed in Table \ref{tab:model_obs}.  The reported values are the combination which had the highest likelihood ${\mathcal L}$, which corresponds to the given $\chi^2$. }
\end{table*}

\section{Discussion}
\label{sec:discussion}
As discussed in Section \ref{sec:pwnmodel}, our modeling of the observed dynamical and spectral properties of \pwn\ allows us to estimate the properties of the associated pulsar -- both its birth properties, the $\gamma$-ray efficiency and spectrum of its magnetospheric emission.
We derived the properties of the pulsar wind powered by the loss of rotational energy of the pulsar, the progenitor supernova explosion, and its environment.  
In Section \ref{sec:disc_pulsar}, we discuss the very high-energy (VHE) emission of the pulsar and compare the observed properties of the magnetospheric $\gamma$-ray emission of this pulsar to that observed from other, rotation-powered, pulsars.  
In Section \ref{sec:disc_model}, we discuss the implication of the derived properties for the supernova explosion.
Lastly, in Section \ref{sec:disc_ic}, we discuss the physical significance concerning our measurements regarding the environment of this source.

\subsection{VHE pulsar emission}
\label{sec:disc_pulsar}
In our analysis of the PWN+PSR complex, at energies above 100\,MeV, we assume a power-law with exponential cutoff for the pulsar. 
We find a good agreement with the observed flux in the 100\,MeV $-$ $\sim$2\,GeV range and conclude it likely that the flux in this energy range is emitted by the pulsar magnetosphere.
In the following sections this magnetospheric emission is compared with its detected MeV emission and to gamma-ray pulsars reported in the second \textit{Fermi} pulsar catalog \citep{2fpc}. 

\subsubsection{Comparison to its MeV pulsed emission}
Since the discovery of pulsar \psr\ in X-rays by \citet{gotthelf2000}, there have been numerous studies related to its high-energy ($\gtrsim$2~keV) pulsed emission. 
\citet{kuiper2009} reported a measurement of the high-energy pulsed spectrum of the pulsar in the $\sim 2-300$\,keV band, and reveiled a major spin-up glitch at the onset of the magnetar-like bursts and enhancement in X-ray flux in 2006 \citep{gavriil2008,kumar2008}.
A first attempt to detect the pulsar at higher energies, in the MeV$-$GeV band of \fermi\, was conducted by \citet{parent2011} using the first $\sim$20 months of data who reported a non-detection of PSR~\psr\ for energies above 100\,MeV.
A follow-up study by \citet{kuiper2018}, who used 8 years (2007 August 28 – 2016 September 4) of {\it Fermi} data and a 
multi-instrument timing solution of the pulsar derived from X-ray observations covering these dates, reported a 4.2$\sigma$ detection of pulsed emission from PSR \psr\ in the $30-100$\,MeV band, with 
no pulsed emission detected at photon energies larger than 100\,MeV, and a pulsed spectrum well described by a log-parabola which peaks at a photon energy of $3.5\pm1.1$\,MeV. 
Our detection at energies above 100\,MeV does not have sufficient photons to allow for a pulsation analysis.

\begin{figure}[ht!]
    \centering
    \includegraphics[width=0.99\columnwidth]{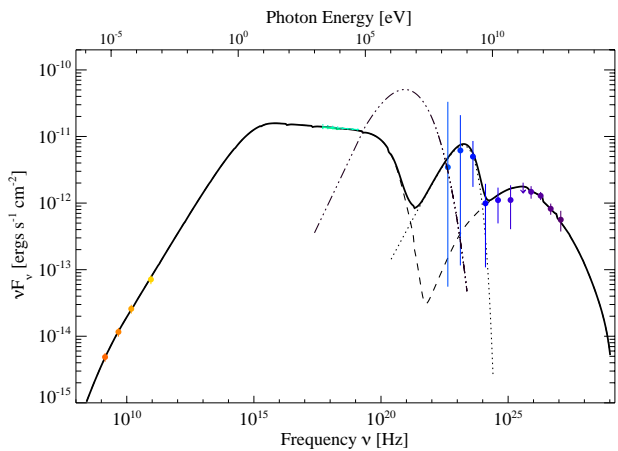}
    \caption{SED of \pwn\ . Observed properties are shown in color and with the best model fit overlain in the black solid line (as shown in figure \ref{fig:flux_sed}). Added is the best-fit pulsed spectrum of PSR \psr\ from \citet{kuiper2018}. } 
    \label{fig:sed_ULcurve}
\end{figure}

Figure \ref{fig:sed_ULcurve} shows our SED model with the pulsar pulsed flux model derived in \citet{kuiper2018} overlaid. 
The flux point in our lowest energy range (centered at 0.176\,GeV, see table \ref{tab:model_obs}) detected in our analysis of the {\it Fermi} data is consistent with this model.
However, the flux measured at higher photon energies is not consistent with an extrapolation of the MeV pulsed emission reported by \citet{kuiper2018}, suggesting the higher energy emission ($>$ 100\,MeV) does not arise from the same emission mechanism.
A similar discrepancy is observed  in the SED of the pulsed emision from the Crab pulsar in this energy, \citep[see Fig. 8,][]{kuiper2018} which also can not be described by a single parabola. 
The here reported GeV component is similar to one predicted for PSR~\psr\ by \citet{harding2017}.
In their modeling of the VHE pulsar emission of \psr, they predict an emission component in the $\sim$\,GeV-range caused by curvature radiation whose predicted shape is qualitatively similar to what is observed here (Figure 3,  \cite{harding2017}).

\subsubsection{Comparison to 2PC}
To determine if the magnetospheric $\gamma$-ray emission of \psr\ is comparable to that of rotation powered pulsars not associated with magnetar-like activities, 
we compare the pulsar properties obtained from our modeling to that of other young, energetic rotation powered pulsars as compiled in the second \textit{Fermi} pulsar catalog \citep[2PC,][]{2fpc}\footnote{Given the detection approach of 2PC, the catalog is biased towards radio-emitting pulsars.}. 
The properties we obtain for the pulsar and compare to the pulsars in \citet{2fpc} are given in Table \ref{tab:model_pars}.
In \citet{2fpc}, Figures 7 to 10 describe pulsar properties from the detected sample that we can compare our obtained properties to.
To magnetic field at the light cylinder is taken from the ATNF catalogue, \citep[$B_{\rm LC} = 1.31\times10^4$\,G,][]{manchester2005}, and the gamma-ray luminosity in the $0.1-100$\,GeV range derived in this work is given in Table \ref{tab:model_pars}.

The gamma-ray efficiency $\eta_{\gamma}$ of $0.97\%$ is typical for gamma-ray pulsars with similar spin-down energy. 
Its gamma-ray luminosity ($L_{\gamma} \approx 7.86\times10^{34}$\,erg\,s$^{-1}$) seems to be on the lower end and more in line with $L_{\gamma} \propto \dot{E}^{\frac{1}{2}}$ than a linear relation, consistent with the findings in 2PC on gamma-ray pulsars.
The obtained photon index of $\Gamma = 1.29$ seems to be harder than most other pulsars of similar spin-down energy, but within the range of values observed in this sample.
The cut-off energy of the spectrum of $\sim 1$\, GeV compared to its magnetic field at the light cylinder is observed to be similar to other pulsars in the $B_{\rm{LC}} \sim 10^4$\,G region.

The obtained pulsar properties agree well with what is obtained for the larger sample of gamma-ray emitting pulsars, supporting the assumption that this additional component originates in the pulsar's magnetosphere, and suggesting that the underlying physical emission mechanism is similar to that observed from rotation-powered pulsars.

\subsection{PWN model}
\label{sec:disc_model}
In \citet{gotthelf2021} we modeled \pwn\ after adding new X-ray measurements as observed in that work.
From our modeling we obtain that the dynamical and spectral properties of \pwn\ can best be described to be formed in a low-energy supernova explosion with low ejecta mass (see Table \ref{tab:model_pars}, $E_{\rm sn} = 1.1\times10^{50}$\, ergs, $M_{\rm ej} = 0.37$\,M$_{\odot}$).
This is very similar, albeit a bit lower, than derived in \citet{gotthelf2021}. 
Other work also seem to favor a low explosion energy and ejecta mass \citep{leahy2008}.
We further compare our results to the modeling results presented in \citet{gotthelf2021}, providing a one on one comparison to the influence of the \fermi\ measurements on the SED and the evolution of the PWN.
From Table \ref{tab:model_pars} it is shown that the addition of the GeV emission shows the need of an additional high-temperature IC photon field.
Next to the additional photon field, this work reports a higher wind magnetization $\eta_{\rm B} \gtrsim 0.04$ than previous studies (see Table \ref{tab:model_pars}). 
This is the result of the increased IC emission required by our \fermi\ detection of this source.  
The additional IC radiative losses associated with this emission decreases the particle energy inside the PWN, and therefore a more strongly magnetized wind is needed to produce the same synchrotron luminiosity.
The increased wind magnetization also affects the low-energy particle index, which is found to be lower in this work.
Furthermore, the slightly longer pulsar spin-down timescale leads to a longer initial spin period.

\subsection{Implications of the additional IC field}
\label{sec:disc_ic}
The energy range observed by \fermi\ provides an excellent view of the ICS part of the spectrum of \pwn\, that was not probed by previous observations performed by the HGPS \citep{HGPS}. 
With the addition of the flux points in the 100\,MeV $-$ 500\, GeV energy range, we need a second background photon field to properly reproduce the IC part of the spectrum in our modeling.

As the case for \citet{gotthelf2021}, one photon field has a temperature of $\sim32~{\rm K}$ which agrees well with the temperature of the surrounding dust \citep{temim2019}.
However, our \fermi\ result requires  an additional, high-temperature photon field, with $T_{\rm ic,2} \sim 1.5\times10^5$\,K and an energy density:
\begin{eqnarray}
\label{eqn:uic2}
u_{\rm ic,2} & = & K_{\rm ic,2} a T_{\rm ic,2}^4 \\
& \approx & 5.4\times10^{-9}~\frac{\rm ergs}{\rm cm^3} = 3.4~\frac{\rm keV}{\rm cm^3},
\end{eqnarray}
where $a$ is the radiation constant, that is orders of magnitude larger than the energy density of the local interstellar radiation field at the ultra-violet (UV) energies where this emission should peak (e.g., \citealt{habing68}).

To identify the possible source of such an intense background photon field, we first calculate the total energy of photons $E_{\rm ic,2}$ inside the PWN, assuming it is a sphere of radius $R_{\rm pwn} = \theta_{\rm pwn}d \approx 0.85~{\rm pc}$ for the values given in Table \ref{tab:model_obs}, such that $E_{\rm ic,2}\sim4\times10^{47}~{\rm ergs}.$  
Assuming that these photons cross the PWN in a time $t_{\rm cross} \sim \frac{2R_{\rm pwn}}{c} \approx 2\times10^8~{\rm s}$, the rate with which such photons are injected into the PWN $L_{\rm ic,2}$ is:
\begin{eqnarray}
\label{eqn:lic2}
L_{\rm ic,2} & \sim & \frac{E_{\rm ic,2}}{t_{\rm cross}} \approx 2.3 \times 10^{39}~\frac{\rm ergs}{\rm s} = 6\times10^5~{\rm L}_\odot.
\end{eqnarray}

Assuming that the energy injection rate $L_{\rm ic,2}$ and temperature $T_{\rm ic,2}$ of this additional background photon field corresponds to the luminosity and (effective) temperature of its source, both values are comparable to that observed from Wolf-Rayet stars (e.g., \citealt{aadland22a, aadland22b, abbott04, tramper15}) -- suggesting the presence of such a star inside the PWN.
If present, this star would have an absolute magnitude $M_V \lesssim -6$ (e.g., \citealt{aadland22a}).  
Using the derived relationship between X-ray absorption and optical extinction (e.g., \cite{foight16}), the interstellar Hydrogen column density towards Kes 75 ($N_H\approx 4\times10^{22}~{\rm cm^{-2}}$; e.g., \citealt{gotthelf2021}) implies an extinction $A_V \gtrsim 14~{\rm mag}$.  
For a distance of $d=5.8~{\rm kpc}$ (Table \ref{tab:model_obs}), the apparent magnitude of a Wolf-Rayet star inside Kes 75 would be $m_V \sim 22$ -- fainter than current surveys of this field (e.g., Pan-STARRS; \citealt{magnier20}).
If present, such a star was likely the binary companion on the stellar progenitor, and would provide important information on its properties and evolution before exploding.

\section{Conclusions}
\label{sec:conclusions}
In this report we analyze the emission associated to the unknown source 4FGL~J1846.9-0247c, coincident with the location of \pwn. 
Performing a detailed analysis of the source points to a PWN+PSR origin from \pwn.
From our modeling of the obtained spectrum, we derive both the physical properties of the PWN, as well as the gamma-ray properties of the pulsar.
The pulsar its gamma-ray parameters are found to be consistent with that observed in pulsed emission from known gamma-ray pulsars.
The magnetospheric component we observe can be explained by a curvature radiation component around these energies as predicted for this source by \cite{harding2017}.

We find the \fermi\ flux measurements provide valuable constraints to the local photon fields, and when combined with the derived progenitor characteristics, support a scenario where the progenitor was in a binary system before going supernova.

The high temperature IC photon field suggests the presence of a Wolf-Rayet star embedded within the PWN. This WR star was then likely the binary companion of the progenitor to \pwn.

Furthermore, the addition of the \fermi\ flux measurements give rise to the need of a higher wind magnetization than perviously reported in \citet{gotthelf2021}, underlining the importance of \fermi\ in the study of PWNe.

\acknowledgments
The authors would like to thank the anonymous referee for their useful comments.
The contributions of JDG and SMS was supported by NASA grant NNX17AL74G issued through the NNH16ZDA001N Astrophysics Data Analysis Program (ADAP).
SMS thanks the \textit{Fermi} summer school for teaching \fermi\ specific analysis.
JDG would like to thank Eric Gotthelf, Samar Safi-Harb, Tea Temim, Maxim Lyutikov, and Jacco Vink for helpful discussions.
We also acknowledge the use of the NASA Astrophysics Data Service (ADS).

\vspace{25mm}

\facilities{Fermi}

\software{
          Fermipy \citep{wood2017},
          ATNF pulsar catalogue \citep{manchester2005}
          }
\bibliography{ms.bbl}

\begin{thebibliography}{}
\expandafter\ifx\csname natexlab\endcsname\relax\def\natexlab#1{#1}\fi
\providecommand{\url}[1]{\href{#1}{#1}}

\bibitem[{{Aadland} {et~al.}(2022{\natexlab{a}}){Aadland}, {Massey}, {Hillier},
  \& {Morrell}}]{aadland22b}
{Aadland}, E., {Massey}, P., {Hillier}, D.~J., \& {Morrell}, N.
  2022{\natexlab{a}}, \apj, 924, 44

\bibitem[{{Aadland} {et~al.}(2022{\natexlab{b}}){Aadland}, {Massey}, {John
  Hillier}, {Morrell}, {Neugent}, \& {Eldridge}}]{aadland22a}
{Aadland}, E., {Massey}, P., {John Hillier}, D., {et~al.} 2022{\natexlab{b}},
  \apj, 931, 157

\bibitem[{{Abbott}(2004)}]{abbott04}
{Abbott}, J.~B. 2004, PhD thesis, University College London, UK

\bibitem[{{Abdo} {et~al.}(2013){Abdo}, {Ajello}, {Allafort}, {Baldini},
  {Ballet}, {Barbiellini}, {Baring}, {Bastieri}, {Belfiore}, {Bellazzini},
  {Bhattacharyya}, {Bissaldi}, {Bloom}, {Bonamente}, {Bottacini}, {Brandt},
  {Bregeon}, {Brigida}, {Bruel}, {Buehler}, {Burgay}, {Burnett}, {Busetto},
  {Buson}, {Caliandro}, {Cameron}, {Camilo}, {Caraveo}, {Casandjian}, {Cecchi},
  {{\c{C}}elik}, {Charles}, {Chaty}, {Chaves}, {Chekhtman}, {Chen}, {Chiang},
  {Chiaro}, {Ciprini}, {Claus}, {Cognard}, {Cohen-Tanugi}, {Cominsky},
  {Conrad}, {Cutini}, {D'Ammando}, {de Angelis}, {DeCesar}, {De Luca}, {den
  Hartog}, {de Palma}, {Dermer}, {Desvignes}, {Digel}, {Di Venere}, {Drell},
  {Drlica-Wagner}, {Dubois}, {Dumora}, {Espinoza}, {Falletti}, {Favuzzi},
  {Ferrara}, {Focke}, {Franckowiak}, {Freire}, {Funk}, {Fusco}, {Gargano},
  {Gasparrini}, {Germani}, {Giglietto}, {Giommi}, {Giordano}, {Giroletti},
  {Glanzman}, {Godfrey}, {Gotthelf}, {Grenier}, {Grondin}, {Grove},
  {Guillemot}, {Guiriec}, {Hadasch}, {Hanabata}, {Harding}, {Hayashida},
  {Hays}, {Hessels}, {Hewitt}, {Hill}, {Horan}, {Hou}, {Hughes}, {Jackson},
  {Janssen}, {Jogler}, {J{\'o}hannesson}, {Johnson}, {Johnson}, {Johnson},
  {Johnson}, {Johnston}, {Kamae}, {Kataoka}, {Keith}, {Kerr}, {Kn{\"o}dlseder},
  {Kramer}, {Kuss}, {Lande}, {Larsson}, {Latronico}, {Lemoine-Goumard},
  {Longo}, {Loparco}, {Lovellette}, {Lubrano}, {Lyne}, {Manchester}, {Marelli},
  {Massaro}, {Mayer}, {Mazziotta}, {McEnery}, {McLaughlin}, {Mehault},
  {Michelson}, {Mignani}, {Mitthumsiri}, {Mizuno}, {Moiseev}, {Monzani},
  {Morselli}, {Moskalenko}, {Murgia}, {Nakamori}, {Nemmen}, {Nuss}, {Ohno},
  {Ohsugi}, {Orienti}, {Orlando}, {Ormes}, {Paneque}, {Panetta}, {Parent},
  {Perkins}, {Pesce-Rollins}, {Pierbattista}, {Piron}, {Pivato}, {Pletsch},
  {Porter}, {Possenti}, {Rain{\`o}}, {Rando}, {Ransom}, {Ray}, {Razzano},
  {Rea}, {Reimer}, {Reimer}, {Renault}, {Reposeur}, {Ritz}, {Romani}, {Roth},
  {Rousseau}, {Roy}, {Ruan}, {Sartori}, {Saz Parkinson}, {Scargle}, {Schulz},
  {Sgr{\`o}}, {Shannon}, {Siskind}, {Smith}, {Spandre}, {Spinelli}, {Stappers},
  {Strong}, {Suson}, {Takahashi}, {Thayer}, {Thayer}, {Theureau}, {Thompson},
  {Thorsett}, {Tibaldo}, {Tibolla}, {Tinivella}, {Torres}, {Tosti}, {Troja},
  {Uchiyama}, {Usher}, {Vandenbroucke}, {Vasileiou}, {Venter}, {Vianello},
  {Vitale}, {Wang}, {Weltevrede}, {Winer}, {Wolff}, {Wood}, {Wood}, {Wood}, \&
  {Yang}}]{2fpc}
{Abdo}, A.~A., {Ajello}, M., {Allafort}, A., {et~al.} 2013, \apjs, 208, 17

\bibitem[{{Abdollahi} {et~al.}(2022){Abdollahi}, {Acero}, {Baldini}, {Ballet},
  {Bastieri}, {Bellazzini}, {Berenji}, {Berretta}, {Bissaldi}, {Blandford},
  {Bloom}, {Bonino}, {Brill}, {Britto}, {Bruel}, {Burnett}, {Buson}, {Cameron},
  {Caputo}, {Caraveo}, {Castro}, {Chaty}, {Cheung}, {Chiaro}, {Cibrario},
  {Ciprini}, {Coronado-Bl{\'a}zquez}, {Crnogorcevic}, {Cutini}, {D'Ammando},
  {De Gaetano}, {Digel}, {Di Lalla}, {Dirirsa}, {Di Venere}, {Dom{\'\i}nguez},
  {Fallah Ramazani}, {Fegan}, {Ferrara}, {Fiori}, {Fleischhack}, {Franckowiak},
  {Fukazawa}, {Funk}, {Fusco}, {Galanti}, {Gammaldi}, {Gargano}, {Garrappa},
  {Gasparrini}, {Giacchino}, {Giglietto}, {Giordano}, {Giroletti}, {Glanzman},
  {Green}, {Grenier}, {Grondin}, {Guillemot}, {Guiriec}, {Gustafsson},
  {Harding}, {Hays}, {Hewitt}, {Horan}, {Hou}, {J{\'o}hannesson}, {Karwin},
  {Kayanoki}, {Kerr}, {Kuss}, {Landriu}, {Larsson}, {Latronico},
  {Lemoine-Goumard}, {Li}, {Liodakis}, {Longo}, {Loparco}, {Lott}, {Lubrano},
  {Maldera}, {Malyshev}, {Manfreda}, {Mart{\'\i}-Devesa}, {Mazziotta}, {Mereu},
  {Meyer}, {Michelson}, {Mirabal}, {Mitthumsiri}, {Mizuno}, {Moiseev},
  {Monzani}, {Morselli}, {Moskalenko}, {Negro}, {Nuss}, {Omodei}, {Orienti},
  {Orlando}, {Paneque}, {Pei}, {Perkins}, {Persic}, {Pesce-Rollins},
  {Petrosian}, {Pillera}, {Poon}, {Porter}, {Principe}, {Rain{\`o}}, {Rando},
  {Rani}, {Razzano}, {Razzaque}, {Reimer}, {Reimer}, {Reposeur},
  {S{\'a}nchez-Conde}, {Saz Parkinson}, {Scotton}, {Serini}, {Sgr{\`o}},
  {Siskind}, {Smith}, {Spandre}, {Spinelli}, {Sueoka}, {Suson}, {Tajima},
  {Tak}, {Thayer}, {Thompson}, {Torres}, {Troja}, {Valverde}, {Wood}, \&
  {Zaharijas}}]{4FGLDR3}
{Abdollahi}, S., {Acero}, F., {Baldini}, L., {et~al.} 2022, \apjs, 260, 53

\bibitem[{{Atwood} {et~al.}(2013){Atwood}, {Albert}, {Baldini}, {Tinivella},
  {Bregeon}, {Pesce-Rollins}, {Sgr{\`o}}, {Bruel}, {Charles}, {Drlica-Wagner},
  {Franckowiak}, {Jogler}, {Rochester}, {Usher}, {Wood}, {Cohen-Tanugi}, \&
  {Zimmer}}]{atwood2013}
{Atwood}, W., {Albert}, A., {Baldini}, L., {et~al.} 2013, arXiv e-prints,
  arXiv:1303.3514

\bibitem[{{Blumer} {et~al.}(2021){Blumer}, {Safi-Harb}, {McLaughlin}, \&
  {Fiore}}]{blumer2021}
{Blumer}, H., {Safi-Harb}, S., {McLaughlin}, M.~A., \& {Fiore}, W. 2021, \apjl,
  911, L6

\bibitem[{{Bock} \& {Gaensler}(2005)}]{bock05}
{Bock}, D.~C.~J., \& {Gaensler}, B.~M. 2005, \apj, 626, 343

\bibitem[{{Bruel} {et~al.}(2018){Bruel}, {Burnett}, {Digel}, {Johannesson},
  {Omodei}, \& {Wood}}]{bruel2018}
{Bruel}, P., {Burnett}, T.~H., {Digel}, S.~W., {et~al.} 2018, arXiv e-prints,
  arXiv:1810.11394

\bibitem[{{Foight} {et~al.}(2016){Foight}, {G{\"u}ver}, {{\"O}zel}, \&
  {Slane}}]{foight16}
{Foight}, D.~R., {G{\"u}ver}, T., {{\"O}zel}, F., \& {Slane}, P.~O. 2016, \apj,
  826, 66

\bibitem[{{Gaensler} \& {Slane}(2006)}]{gaensler06}
{Gaensler}, B.~M., \& {Slane}, P.~O. 2006, \araa, 44, 17

\bibitem[{{Gavriil} {et~al.}(2008){Gavriil}, {Gonzalez}, {Gotthelf}, {Kaspi},
  {Livingstone}, \& {Woods}}]{gavriil2008}
{Gavriil}, F.~P., {Gonzalez}, M.~E., {Gotthelf}, E.~V., {et~al.} 2008, Science,
  319, 1802

\bibitem[{{Gelfand} {et~al.}(2013){Gelfand}, {Slane}, \& {Temim}}]{gelfand13}
{Gelfand}, J., {Slane}, P., \& {Temim}, T. 2013, in The Fast and the Furious:
  Energetic Phenomena in Isolated Neutron Stars, Pulsar Wind Nebulae and
  Supernova Remnants, ed. J.~U. {Ness}, 24

\bibitem[{{Gelfand} {et~al.}(2014){Gelfand}, {Slane}, \& {Temim}}]{gelfand2014}
{Gelfand}, J.~D., {Slane}, P.~O., \& {Temim}, T. 2014, Astronomische
  Nachrichten, 335, 318

\bibitem[{{Gelfand} {et~al.}(2015){Gelfand}, {Slane}, \& {Temim}}]{gelfand15}
---. 2015, \apj, 807, 30

\bibitem[{{Gelfand} {et~al.}(2009){Gelfand}, {Slane}, \& {Zhang}}]{gelfand09}
{Gelfand}, J.~D., {Slane}, P.~O., \& {Zhang}, W. 2009, \apj, 703, 2051

\bibitem[{{Gotthelf} {et~al.}(2021){Gotthelf}, {Safi-Harb}, {Straal}, \&
  {Gelfand}}]{gotthelf2021}
{Gotthelf}, E.~V., {Safi-Harb}, S., {Straal}, S.~M., \& {Gelfand}, J.~D. 2021,
  \apj, 908, 212

\bibitem[{{Gotthelf} {et~al.}(2000){Gotthelf}, {Vasisht}, {Boylan-Kolchin}, \&
  {Torii}}]{gotthelf2000}
{Gotthelf}, E.~V., {Vasisht}, G., {Boylan-Kolchin}, M., \& {Torii}, K. 2000,
  \apjl, 542, L37

\bibitem[{{Gotthelf} {et~al.}(2014){Gotthelf}, {Tomsick}, {Halpern}, {Gelfand
  }, {Harrison}, {Boggs}, {Christensen}, {Craig}, {Hailey}, {Kaspi}, {Stern},
  \& {Zhang}}]{gotthelf14}
{Gotthelf}, E.~V., {Tomsick}, J.~A., {Halpern}, J.~P., {et~al.} 2014, \apj,
  788, 155

\bibitem[{{Gull{\'o}n} {et~al.}(2015){Gull{\'o}n}, {Pons}, {Miralles},
  {Vigan{\`o}}, {Rea}, \& {Perna}}]{gullon15}
{Gull{\'o}n}, M., {Pons}, J.~A., {Miralles}, J.~A., {et~al.} 2015, \mnras, 454,
  615

\bibitem[{{H.~E.~S.~S. Collaboration} {et~al.}(2018){H.~E.~S.~S.
  Collaboration}, {Abdalla}, {Abramowski}, {Aharonian}, {Ait Benkhali},
  {Ang{\"u}ner}, {Arakawa}, {Arrieta}, {Aubert}, {Backes}, {Balzer}, {Barnard},
  {Becherini}, {Becker Tjus}, {Berge}, {Bernhard}, {Bernl{\"o}hr}, {Blackwell},
  {B{\"o}ttcher}, {Boisson}, {Bolmont}, {Bonnefoy}, {Bordas}, {Bregeon},
  {Brun}, {Brun}, {Bryan}, {B{\"u}chele}, {Bulik}, {Capasso}, {Carrigan},
  {Caroff}, {Carosi}, {Casanova}, {Cerruti}, {Chakraborty}, {Chaves}, {Chen},
  {Chevalier}, {Colafrancesco}, {Condon}, {Conrad}, {Davids}, {Decock}, {Deil},
  {Devin}, {deWilt}, {Dirson}, {Djannati-Ata{\"\i}}, {Domainko}, {Donath},
  {Drury}, {Dutson}, {Dyks}, {Edwards}, {Egberts}, {Eger}, {Emery},
  {Ernenwein}, {Eschbach}, {Farnier}, {Fegan}, {Fernand es}, {Fiasson},
  {Fontaine}, {F{\"o}rster}, {Funk}, {F{\"u}{\ss}ling}, {Gabici}, {Gallant},
  {Garrigoux}, {Gast}, {Gat{\'e}}, {Giavitto}, {Giebels}, {Glawion},
  {Glicenstein}, {Gottschall}, {Grondin}, {Hahn}, {Haupt}, {Hawkes},
  {Heinzelmann}, {Henri}, {Hermann}, {Hinton}, {Hofmann}, {Hoischen}, {Holch},
  {Holler}, {Horns}, {Ivascenko}, {Iwasaki}, {Jacholkowska}, {Jamrozy},
  {Jankowsky}, {Jankowsky}, {Jingo}, {Jouvin}, {Jung-Richardt}, {Kastendieck},
  {Katarzy{\'n}ski}, {Katsuragawa}, {Katz}, {Kerszberg}, {Khangulyan},
  {Kh{\'e}lifi}, {King}, {Klepser}, {Klochkov}, {Klu{\'z}niak}, {Komin},
  {Kosack}, {Krakau}, {Kraus}, {Kr{\"u}ger}, {Laffon}, {Lamanna}, {Lau},
  {Lees}, {Lefaucheur}, {Lemi{\`e}re}, {Lemoine-Goumard}, {Lenain}, {Leser},
  {Lohse}, {Lorentz}, {Liu}, {L{\'o}pez-Coto}, {Lypova}, {Marandon},
  {Malyshev}, {Marcowith}, {Mariaud}, {Marx}, {Maurin}, {Maxted}, {Mayer},
  {Meintjes}, {Meyer}, {Mitchell}, {Moderski}, {Mohamed}, {Mohrmann},
  {Mor{\r{a}}}, {Moulin}, {Murach}, {Nakashima}, {de Naurois}, {Ndiyavala},
  {Niederwanger}, {Niemiec}, {Oakes}, {O'Brien}, {Odaka}, {Ohm}, {Ostrowski},
  {Oya}, {Padovani}, {Panter}, {Parsons}, {Paz Arribas}, {Pekeur}, {Pelletier},
  {Perennes}, {Petrucci}, {Peyaud}, {Piel}, {Pita}, {Poireau}, {Poon},
  {Prokhorov}, {Prokoph}, {P{\"u}hlhofer}, {Punch}, {Quirrenbach}, {Raab},
  {Rauth}, {Reimer}, {Reimer}, {Renaud}, {de los Reyes}, {Rieger}, {Rinchiuso},
  {Romoli}, {Rowell}, {Rudak}, {Rulten}, {Safi-Harb}, {Sahakian}, {Saito},
  {Sanchez}, {Santangelo}, {Sasaki}, {Schand ri}, {Schlickeiser},
  {Sch{\"u}ssler}, {Schulz}, {Schwanke}, {Schwemmer}, {Seglar-Arroyo},
  {Settimo}, {Seyffert}, {Shafi}, {Shilon}, {Shiningayamwe}, {Simoni}, {Sol},
  {Spanier}, {Spir-Jacob}, {Stawarz}, {Steenkamp}, {Stegmann}, {Steppa},
  {Sushch}, {Takahashi}, {Tavernet}, {Tavernier}, {Taylor}, {Terrier},
  {Tibaldo}, {Tiziani}, {Tluczykont}, {Trichard}, {Tsirou}, {Tsuji}, {Tuffs},
  {Uchiyama}, {van der Walt}, {van Eldik}, {van Rensburg}, {van Soelen},
  {Vasileiadis}, {Veh}, {Venter}, {Viana}, {Vincent}, {Vink}, {Voisin},
  {V{\"o}lk}, {Vuillaume}, {Wadiasingh}, {Wagner}, {Wagner}, {Wagner}, {White},
  {Wierzcholska}, {Willmann}, {W{\"o}rnlein}, {Wouters}, {Yang}, {Zaborov},
  {Zacharias}, {Zanin}, {Zdziarski}, {Zech}, {Zefi}, {Ziegler}, {Zorn}, \&
  {{\.Z}ywucka}}]{HGPS}
{H.~E.~S.~S. Collaboration}, {Abdalla}, H., {Abramowski}, A., {et~al.} 2018,
  \aap, 612, A1

\bibitem[{{Habing}(1968)}]{habing68}
{Habing}, H.~J. 1968, \bain, 19, 421

\bibitem[{{Harding}(2013)}]{harding2013}
{Harding}, A.~K. 2013, Frontiers of Physics, 8, 679

\bibitem[{{Harding} \& {Kalapotharakos}(2017)}]{harding2017}
{Harding}, A.~K., \& {Kalapotharakos}, C. 2017, in Proceedings of the 7th
  International Fermi Symposium, 6

\bibitem[{{Hattori} {et~al.}(2020){Hattori}, {Straal}, {Zhang}, {Temim},
  {Gelfand}, \& {Slane}}]{hattori2020}
{Hattori}, S., {Straal}, S.~M., {Zhang}, E., {et~al.} 2020, \apj, 904, 32

\bibitem[{{Kaspi}(2018)}]{kaspi2018}
{Kaspi}, V.~M. 2018, in Pulsar Astrophysics the Next Fifty Years, ed.
  P.~{Weltevrede}, B.~B.~P. {Perera}, L.~L. {Preston}, \& S.~{Sanidas}, Vol.
  337, 3--8

\bibitem[{{Kaspi} \& {Beloborodov}(2017)}]{kaspi2017}
{Kaspi}, V.~M., \& {Beloborodov}, A.~M. 2017, \araa, 55, 261

\bibitem[{{Krimm} {et~al.}(2020){Krimm}, {Lien}, {Page}, {Palmer},
  {Tohuvavohu}, \& {Neil Gehrels Swift Observatory Team}}]{krimm2020}
{Krimm}, H.~A., {Lien}, A.~Y., {Page}, K.~L., {et~al.} 2020, GRB Coordinates
  Network, 28187, 1

\bibitem[{{Kuiper} \& {Hermsen}(2009)}]{kuiper2009}
{Kuiper}, L., \& {Hermsen}, W. 2009, \aap, 501, 1031

\bibitem[{{Kuiper} {et~al.}(2018){Kuiper}, {Hermsen}, \& {Dekker}}]{kuiper2018}
{Kuiper}, L., {Hermsen}, W., \& {Dekker}, A. 2018, \mnras, 475, 1238

\bibitem[{{Kumar} \& {Safi-Harb}(2008)}]{kumar2008}
{Kumar}, H.~S., \& {Safi-Harb}, S. 2008, \apjl, 678, L43

\bibitem[{{Leahy} \& {Tian}(2008)}]{leahy2008}
{Leahy}, D.~A., \& {Tian}, W.~W. 2008, \aap, 480, L25

\bibitem[{{Livingstone} {et~al.}(2011){Livingstone}, {Ng}, {Kaspi}, {Gavriil},
  \& {Gotthelf}}]{livingstone11}
{Livingstone}, M.~A., {Ng}, C.~Y., {Kaspi}, V.~M., {Gavriil}, F.~P., \&
  {Gotthelf}, E.~V. 2011, \apj, 730, 66

\bibitem[{{Magnier} {et~al.}(2020){Magnier}, {Sweeney}, {Chambers},
  {Flewelling}, {Huber}, {Price}, {Waters}, {Denneau}, {Draper}, {Farrow},
  {Jedicke}, {Hodapp}, {Kaiser}, {Kudritzki}, {Metcalfe}, {Stubbs}, \&
  {Wainscoat}}]{magnier20}
{Magnier}, E.~A., {Sweeney}, W.~E., {Chambers}, K.~C., {et~al.} 2020, \apjs,
  251, 5

\bibitem[{{Manchester} {et~al.}(2005){Manchester}, {Hobbs}, {Teoh}, \&
  {Hobbs}}]{manchester2005}
{Manchester}, R.~N., {Hobbs}, G.~B., {Teoh}, A., \& {Hobbs}, M. 2005, \aj, 129,
  1993

\bibitem[{{Ng} {et~al.}(2008){Ng}, {Slane}, {Gaensler}, \& {Hughes}}]{ng2008}
{Ng}, C.~Y., {Slane}, P.~O., {Gaensler}, B.~M., \& {Hughes}, J.~P. 2008, \apj,
  686, 508

\bibitem[{{Parent} {et~al.}(2011){Parent}, {Kerr}, {den Hartog}, {Baring},
  {DeCesar}, {Espinoza}, {Gotthelf}, {Harding}, {Johnston}, {Kaspi},
  {Livingstone}, {Romani}, {Stappers}, {Watters}, {Weltevrede}, {Abdo},
  {Burgay}, {Camilo}, {Craig}, {Freire}, {Giordano}, {Guillemot}, {Hobbs},
  {Keith}, {Kramer}, {Lyne}, {Manchester}, {Noutsos}, {Possenti}, \&
  {Smith}}]{parent2011}
{Parent}, D., {Kerr}, M., {den Hartog}, P.~R., {et~al.} 2011, \apj, 743, 170

\bibitem[{{Reynolds} {et~al.}(2018){Reynolds}, {Borkowski}, \&
  {Gwynne}}]{reynolds18}
{Reynolds}, S.~P., {Borkowski}, K.~J., \& {Gwynne}, P.~H. 2018, \apj, 856, 133

\bibitem[{{Salter} {et~al.}(1989){Salter}, {Reynolds}, {Hogg}, {Payne}, \&
  {Rhodes}}]{salter89a}
{Salter}, C.~J., {Reynolds}, S.~P., {Hogg}, D.~E., {Payne}, J.~M., \& {Rhodes},
  P.~J. 1989, \apj, 338, 171

\bibitem[{{Temim} {et~al.}(2019){Temim}, {Slane}, {Sukhbold}, {Koo}, {Raymond},
  \& {Gelfand}}]{temim2019}
{Temim}, T., {Slane}, P., {Sukhbold}, T., {et~al.} 2019, \apjl, 878, L19

\bibitem[{{Tramper} {et~al.}(2015){Tramper}, {Straal}, {Sanyal}, {Sana}, {de
  Koter}, {Gr{\"a}fener}, {Langer}, {Vink}, {de Mink}, \& {Kaper}}]{tramper15}
{Tramper}, F., {Straal}, S.~M., {Sanyal}, D., {et~al.} 2015, \aap, 581, A110

\bibitem[{{Verbiest} {et~al.}(2012){Verbiest}, {Weisberg}, {Chael}, {Lee}, \&
  {Lorimer}}]{verbiest2012}
{Verbiest}, J.~P.~W., {Weisberg}, J.~M., {Chael}, A.~A., {Lee}, K.~J., \&
  {Lorimer}, D.~R. 2012, \apj, 755, 39

\bibitem[{{Wood} {et~al.}(2017){Wood}, {Caputo}, {Charles}, {Di Mauro},
  {Magill}, {Perkins}, \& {Fermi-LAT Collaboration}}]{wood2017}
{Wood}, M., {Caputo}, R., {Charles}, E., {et~al.} 2017, in International Cosmic
  Ray Conference, Vol. 301, 35th International Cosmic Ray Conference
  (ICRC2017), 824

\end{thebibliography}
\bibliographystyle{aasjournal}
\end{document}